\documentclass[pre,preprint,showpacs,preprintnumbers,amsmath,amssymb]{revtex4}

\usepackage{graphicx}
\usepackage{dcolumn}
\usepackage{bm}
\usepackage[mathscr]{eucal}
\usepackage{mathrsfs}

\newtheorem{Definition}{Definition}
\newtheorem{Corollary}{Corollary}

\begin{document}
\title{Nature of the order parameter of glass}

\author{Koun Shirai$^{1,2}$ }
\affiliation{%
$^{1}$Vietnam Japan University, Vietnam National University, Hanoi, \\
Luu Huu Phuoc Road, My Dinh 1 Ward, Nam Tu Liem District, Hanoi, Vietnam \\
%
%
$^{2}$SANKEN, Osaka University, 
8-1 Mihogaoka, Ibaraki, Osaka 567-0047, Japan
}%


\begin{abstract}
In glass physics, order parameters have long been used in the thermodynamic description of glasses, but the nature is not yet clear. The difficulty is how to find order in disordered systems. 
The conventional treatments of glass cause serious internal inconsistencies in thermodynamics.
The issue turns out to be ascribed to a fundamental problem of defining state variables in thermodynamics. This study provides a coherent understanding of the nature of order parameters for glasses and crystals, starting from the fundamental issue of defining state variables in thermodynamics.
The state variable is defined as the time-averaged value of a dynamical variable under the constraint conditions, when equilibrium is established. It gives the same value at any time it is measured as far as the equilibrium state is maintained. This constancy is here called time invariance. 
From this definition, it is deduced that the state variables of a solid are the time-averaged positions $\{ \bar{\bf R}_{j} \}$ of all atoms constituting the solid, regardless of periodicity. 
The general properties of order parameters are fully compatible with the time invariance of state variables. For glass, $\{ \bar{\bf R}_{j} \}$ are state variables and order parameters too. Because $\{ \bar{\bf R}_{j} \}$ are order parameters, any property determined by $\{ \bar{\bf R}_{j} \}$ can also be an order parameter. This explains why many quantities can be used for order parameters. Symmetry introduces redundancy among the averaged positions of the atoms, reducing the number of independent state variables. The often-claimed association of order to periodicity lacks generality as shown by abundant examples of well-ordered material, such as DNA. Order in thermodynamics means the time invariance.
\end{abstract}


\maketitle


\section{Introduction}

\subsection{Historical background}

In glass physics, since Prigogine and Defay introduced the idea \cite{Berthier11,Berthier16,Prigogine54,Davies53a,Davies53, Nemilov-VitreousState}, order parameters have been indispensable for theoretical studies in the glass transition. 
Despite their importance, the nature of order parameters of glass has not been clarified. 
For crystals, the meaning of the order parameter is intuitively clear. It is a quantity indicating a long-range ordered state, such as magnetization for ferromagnetic materials \cite{Blundell01,Sethna}. 
Landau originally introduced the notion to indicate the emergency of symmetry in phase transitions (\cite{Landau-SP}, chap.~14).
From this original meaning, it is no sense to speak of order for disordered systems. 
Thus physicists have attempted to find physical meaning in the order parameter of glass. Naturally, several ways have been proposed to define the order parameter. They extend from geometrical parameters, such as four-point correlation function \cite{Berthier11}, polyhedron order \cite{Xia15}, and orientational order \cite{Yang19}, to more abstract ones, such as the overlap function or caging order parameter \cite{Parisi83,Franz97,Charbonneau14}, which are derivatives from Edwards-Anderson order parameter in spin glass \cite{note1}.
In the following, we summarize a general (but not necessarily universal) understanding of the order parameter of glass. In this paper, the order parameter (OP) is denoted by $z$. 

\paragraph{Kinetic parameter.} 
Since a chemical reaction is a process to approach an equilibrium state, $z$ is considered a kinetic parameter called advancement indicating the degree to which a chemical reaction proceeds \cite{Prigogine54}. Since the glass transition is the process of approaching the glass state, this analogy makes sense. 
In general, $z$ is not considered a state variable in a chemical reaction because a reaction is a nonequilibrium process. However, by introducing affinity $A$, the kinetic parameter $z$ is coped in the standard formulation of thermodynamics, as
\begin{equation}
dG = -SdT +Vdp -Adz,
\label{eq:UandAdz}
\end{equation}
where $G$ is Gibbs free energy, $S$ is entropy, $p$ is pressure, and $V$ is volume. Affinity is the driving force for a reaction and it is a function of state variables $T$ and $p$ plus $z$, as $A=A(T, p, z)$. With appropriate scaling, $z$ can vary from 1 to 0, with 0 corresponding to equilibrium. As the system approaches equilibrium, $A$ vanishes. Hence, no kinetic parameters remain at equilibrium, as should be. 
However, equilibrium is never established in the glass transition because of its large and slow retardation of atom rearrangement. Structural changes end at a finite value $z_{0} \neq 0$ below the transition temperature $T_{g}$. This glass state is called the frozen state.
The value $z_{0}$ is subject to the energy-minimum condition
\begin{equation}
\left( \frac{\partial G}{\partial z} \right)_{T,p}=0, 
\label{eq:dgdz-zero}
\end{equation}
which, in turn, gives the dependence of $z_{0}$ on $T$ and $p$, as $z_{0}=z_{0}(T,p)$. Accordingly, $z_{0}$ of the frozen state is not an independent variable. Although it is originally a kinetic parameter, it becomes a state variable in the frozen state \cite{Prigogine99,Nemilov-VitreousState}. 
This is not a contradiction. The difference in the character of the state and dynamical variables of the frozen state is tuned based on local equilibrium Ansatz \cite{Prigogine99}. The problem in the case of glass is that it is difficult to give a specific interpretation to ``the degree of advancement" for the glass transition. This is in contrast to the case of chemical reactions, where $z$ has a clear meaning of a change in the components of molecular species. The difficulty increases when one recognizes that there are multiple OPs (see Sec.~\ref{sec:Plurality}).

\paragraph{Structural parameter.}  
At this point, we have to declare the scope of this study. The main interest of this study is not the process but the material property. Therefore, we restrict our argument to OP as material property. This only focuses on the state variable $z_{0}$ of the frozen state.
Liquid states are structureless. Instead, they have a large number of atom configurations. This freedom in configuration is reduced at the glass transition. Hence, it is natural to consider that this reduction in configurational freedom indicates the occurrence of structural fractions. By observing the short-range order of the radial distribution function (RDF), this short-range order deserves a glass OP \cite{Davies53a,Elliott90}. However, the differences observed in RDF between glass and liquid are not sharp. On account of the drastic change in atom mobility at the glass transition, it is unsuitable to characterize the glass transition with such a dull quantity. The obvious difference between glass and liquid is atom mobility: atoms are mobile in liquids but not in solids. This aspect can be presented by time-correlation function. Thus, the dynamic density-density correlation functions become essential tools for studying the glass transition \cite{Radons-book05}. On the other hand, dynamic correlation functions do not provide more accurate structural information than RDF does. They focus on the time dependence, $z(t)$, but do not resolve the detailed dependence on $z_{0}$ of various frozen states.

The difficulty is that the same kind of glass exhibits different properties depending on the thermal history, although their atomic arrangements look similar \cite{Berthier19}. For this reason, glass manufacturers particularly focus on preparation conditions.
RDF is not sensitive to the structural details of the frozen state. Beyond RDF, no unique way exists to define ``higher-order" OPs. With this difficulty, researchers sometimes refer to it as hidden order \cite{Binder86,Wang-Z14,Xia15,Tong18,Berthier19a,Tanaka22}, a term that gives a somewhat mysterious tone to the glass OP.
The problem is essentially ascribed to how the glass structure is identified.
Consider two samples of the same glass. Despite the same chemical composition, generally they have different atomic arrangements. The question is whether the two samples have the same OPs. Even it may be unclear what the ``same" glass means \cite{Shirai18-StateVariable}.

\paragraph{Hysteresis problem.} The last issue raises the entangling problem between the properties of glass and the process. It is known that the current properties of a glass are affected by its history, including the cooling rate at which it was obtained. This behavior is known as hysteresis. Previously, we said that the glass OP, $z_{0}$, is a state variable, and $z_{0}$ is controlled only by $T$ and $p$. We have learned that the current properties of glass are not uniquely determined by current $T$ and $p$ alone. It is really a contradiction.
When hysteresis appears in a material property, the material is generally considered to be in a nonequilibrium state. Describing hysteresis thermodynamically in terms of state variables has long been a conundrum since it was first addressed by Bridgman \cite{Bridgman50}. The thermodynamic description of hysteresis by adding internal variables to usual state variables has been actively studied \cite{Coleman67,Rice71,Muschik93,Maugin93,Maugin94,Restuccia88}. (We use internal variables and OP interchangeably here.)
Although this approach may succeed to some extent, the nature of the internal variable also is not known. Because of this difficulty, some authors argue that ``internal variables can eventually be measured by a {\em gifted} experimentalist but cannot be controlled" (in Ref.~\cite{Maugin93}, p.~224).

The serious problem in the conventional view that hysteresis is a nonequilibrium property is that it leads to destructive consequences for thermodynamics \cite{Kestin70,Kestin92}. 
Close observations reveal that hysteresis is a property that any solid has to some degree \cite{Shirai18-StateVariable}. The mechanical properties of metals depend on their thermal treatment history. For materials used for memory, the current state is determined by a field previously applied. The electrical properties of semiconductors are affected by the preparation conditions due to the presence of intrinsic defects and dislocations. Viewing glass as nonequilibrium state merely because of hysteresis leads to the unpleasant conclusion that all solids are in a nonequilibrium state. No one has solved this difficulty until now.

\paragraph{Central contention.}
In the last argument, we noticed that the OP problem of glass is part of a general problem that seriously affects our understanding of the thermodynamics of solids. 
By focusing on the arguments from (a) to (c), we can figure out that the serious difficulty (all solids turn to be nonequilibrium) is based on the understanding that the equilibrium properties of a solid are a function of $T$ and $p$ (or $V$) only. This is true in the gaseous state, but there is no rigorous proof for solids. The problem eventually goes back to the fundamental problem what an equilibrium is for solids \cite{Shirai18-StateVariable}. Without revisiting the fundamental problem of thermodynamics, the elusive notion of glass OP cannot be understood. 

Another aspect of the OP problem in glass is the notion of disorder in thermodynamics. In solid state physics, order means periodicity. However, thermodynamics is a closed theory and does not need the help of other disciplines. Therefore, the theory should not depend on the details of the internal structure of the system. The term disorder in thermodynamics is most often used to describe entropy. Traditionally, entropy has been interpreted as ``the degree of disorder". However, a more appropriate interpretation is {\it missing information} \cite{Ben-Naim} (see Sec.~\ref{sec:randomness}). From this perspective, the meaning of order for glass is investigated here.

\subsection{Purpose and strategy}
This paper presents a coherent view of glass OP from fundamental laws of thermodynamics.
The glass problem is not treated as an exception, but as part of a general theory. The conclusion is that the OP of glass as well as other solids are equivalent to state variables in thermodynamics. The state variables in a solid are the equilibrium positions of atoms in the solid. 
To draw this conclusion, we begin our argument with the definition of equilibrium (Sec.~\ref{sec:equilibrium}). Next, we explain how the state variables are defined and show the results for solids (Sec.~\ref{sec:constraint}). Based on this, we find equivalence between the OP and the state variables. 
The arguments in Secs.~\ref{sec:equilibrium} and \ref{sec:constraint} were presented in an early work \cite{Shirai18-StateVariable} but are included here because of completeness of theory. Finally, we discuss the key aspects of OP in the glass issue (Sec.~\ref{sec:discussion}).

\paragraph{Scope.}
Before proceeding to the main theory, we must set a limit because the topics are too broad to treat in a single paper.
The question is, what a thermodynamic approach means for those phenomena which are traditionally considered nonequilibrium?
\begin{figure}[htbp]
    \centering
    \includegraphics[width=100 mm, bb=0 0 420 280]{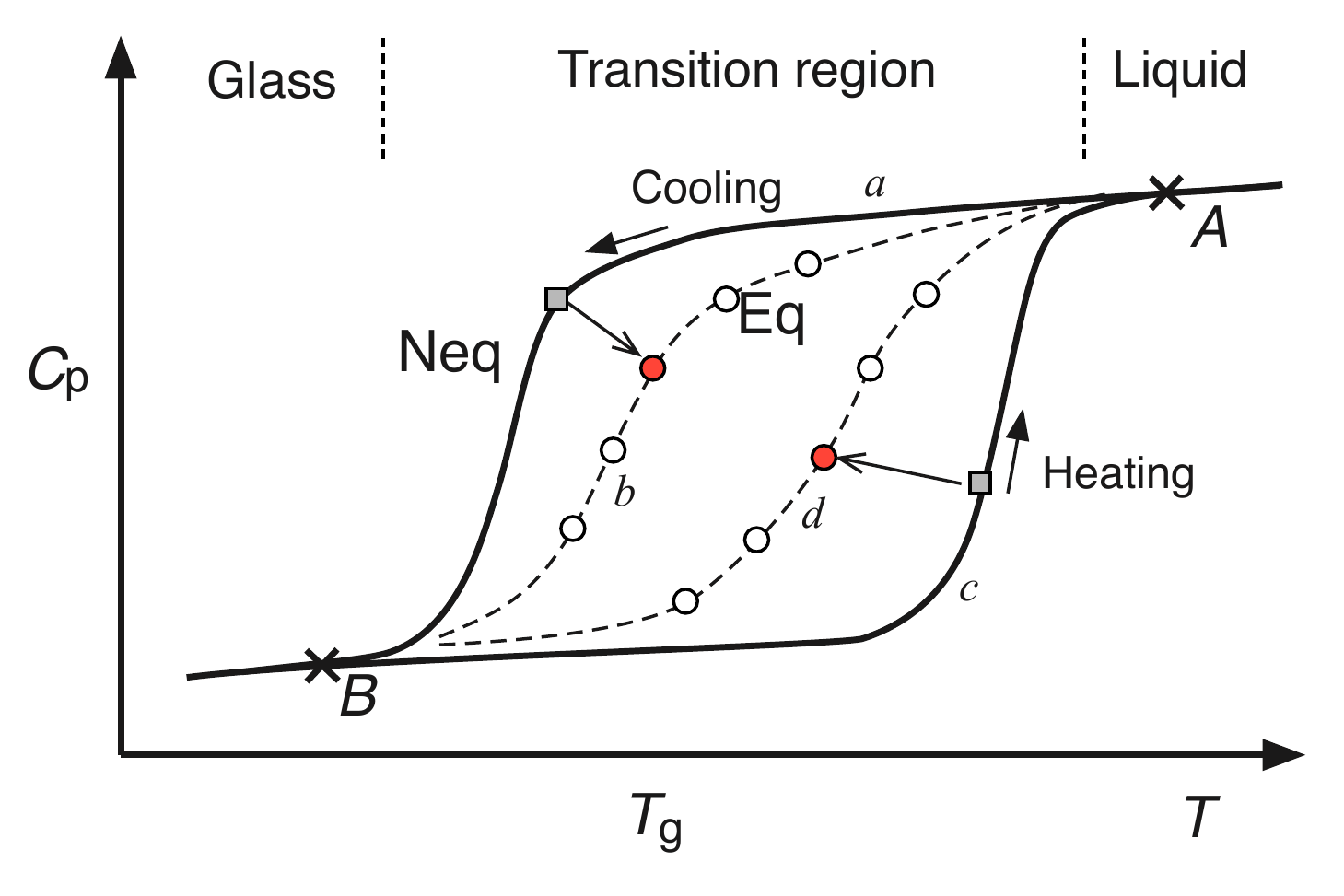} 
  \caption{Specific heat $C_{p}$ of glass at the glass transition. The transition region is defined by the region where $C_{p}$ rapidly changes. The glass transition temperature $T_{g}$ is roughly given by the middle point of the region. Solid curves indicate the actual cooling or heating processes. Dashed curves indicate the equilibrium curves obtained from the nonequilibrium states in adiabatic relaxation processes.
  } \label{fig:CTexperiment}
\end{figure}
Since a process represents changes in state over time, the state in the process must, by definition, be nonequilibrium. At this point, there is no point in discussing it. Figure \ref{fig:CTexperiment} shows the specific heat $C_{p}$ of glass around the glass transition temperature $T_{g}$ in the cooling or heating processes. This exhibits hysteresis between the cooling and heating processes. 
When glass is cooled from liquid state $A$ at a constant rate, a $C_{p}-T$ curve like a solid line $a$ is obtained. The states on this line are nonequilibrium because they have time dependence. Now, let us stop the temperature scan at $T$ (gray squares) and quickly isolate the sample from the heat bath. The glass will then undergo an adiabatic relaxation and eventually reach a state that is no longer to change. This static state (open circles) is called equilibrium in the present context. The important question here is whether this is an equilibrium state. The definition of equilibrium is the subject of this paper, which is discussed in Sec.~\ref{sec:equilibrium}. Repeating this relaxation process at various temperatures on the original nonequilibrium curve $a$ yields a continuous equilibrium curve $b$ (dashed lines).

The thermal path $b$ thus obtained is called a quasi-static process in the thermodynamics context. This study focuses on the static states on this quasi-static path. In usual sense of thermodynamics, all states in a quasi-static process must be in equilibrium. The big question is how they are thermodynamically described. The question is answered in Sec.~\ref{sec:constraint}. By restricting the OP of a glass to those static states only, we can construct theory in a clear manner. Any thermodynamic function, such as entropy, can be obtained by integrating $C_{p}$ along the path $b$. 
Note that this restriction on the thermal path does not necessarily mean that the final state $B$ of the quasi-static cooling process is unique. Many different quasi-static paths $b'$ starting from the same state $A$ exist, ending at different equilibrium states $B'$.

\section{Theory}
\label{sec:theory}

\subsection{Thermodynamic equilibrium}
\label{sec:equilibrium}

Classical thermodynamics consists of four laws, from the zeroth to the third Law. The zeroth law describes a property of thermodynamic equilibrium. It is a transitive property: if the systems $A$ and $B$, and independently $A$ and $C$, are in equilibrium, the systems $B$ and $C$ will also be in equilibrium. Bringing a block of 50$^{\circ}$C copper into contact with a block of 50$^{\circ}$C aluminum does not cause any change. Glass is no exception. No changes occur when a piece of glass at 50$^{\circ}$C and a block of aluminum at 50$^{\circ}$C are brought into contact. 
Although the transitive property is a property of equilibrium, it is not a sufficient condition for equilibrium. Further conditions are required that the other state variables also remain unchanged. The zeroth law provides the definition of the empirical temperature as a universal measure of equilibrium but does not provide the definition of equilibrium.

One may define equilibrium as the state that all the state variables do not change. This definition naturally raises another question: which are state variables? Obviously, temperature and pressure are the answer for gases. However, are there any other variables independent of $T$ and $p$? (In this study, chemical reactions are not treated, and hence mole fraction is excluded.) This question is by no means trivial for solids. Callen posed this question, in his textbook, when the definition of equilibrium is discussed (\cite{Callen}, Sec.~1.5). He noticed that, for many solids, the current properties are affected by the past history. Plastically deformed solids cannot be described by $T$ and $V$ solely \cite{Berdichevsky06}. Callen finally failed to answer the question whether the current properties can be specified by adding appropriate state variables. Nobody has succeeded.
The difficulty is that the word state variable is already used to define equilibrium, but the state variable presupposes equilibrium. The argument falls into an egg and chicken problem. This dilemma has been recently solved by Gyftopoulos and Berreta \cite{Gyftopoulos}. They use only the second law for defining equilibrium. No law is more fundamental than the second law, so that the derived definition is quite general. From the second law, we know the property of equilibrium as
\begin{Definition}{\rm (Equilibrium)}
it is impossible to extract work from a system in equilibrium to the outside without leaving any effects on the environment. {\rm (\cite{Gyftopoulos}, p.~58)}  %
\label{def:equilibrium}
\end{Definition} 
We can now proceed in the reverse order and use this as the definition of equilibrium. No use is made of the word state variable, and thus no self-contradiction occurs. Definition 1 is consistent with our understanding for the equilibrium character of existing solids. At this point, glasses are no exception. Glasses must be in equilibrium, because work cannot be extracted from them.

Definition 1 says that work can be obtained only when a system undergoes spontaneous change. 
In the glass literature, it is often claimed that glass changes slowly toward equilibrium. Since this is a spontaneous change, readers might argue that work can be obtained in that period. However, in the real world, there is no eternal equilibrium. Such a sluggish change commonly occurs. The state of the gas in a gas cylinder is equilibrium. However, even a robust metal wall is subject to gas leakage, and the gas will eventually escape from the cylinder. A mixture of nitrogen and hydrogen molecules is chemically inactive at normal temperatures. However, the mixture will react to produce ammonia gas in the age of Jupiter: the free energy of the latter is lower than the former. A metal block is in equilibrium but will be oxidized for years. Although diamond does not change under normal conditions, an energetic investigation shows that it will eventually be transformed into graphite. Considering nuclear reactions into account, only iron should be in stable equilibrium.
These observations indicate that equilibrium must be argued under certain restrictions. Namely, when changes in property can be ignored within a certain time, we can argue equilibrium \cite{note2}. 
The characteristic time is the relaxation time of that property.

\subsection{Constraints and state variables}
\label{sec:constraint}
What factors determine the relaxation time? This problem was addressed by Gibbs in his classical book more than a century \cite{Gibbs}.
He investigated the distinction between equilibrium and static states. Equilibrium is a static state sustained by two competitive active tendencies that work in opposite directions. There is also another type of static states. For example, a mixture of nitrogen and hydrogen gases is stable at normal conditions, maintaining a static state. However, this static state varies at high temperatures, so they can react to produce ammonia. Gibbs called it passive resistance, which prevents them from reacting. Since then, different names have been used in different areas, such as inhibitor and anti-catalysis \cite{Hatsopoulos}. In today's context of thermodynamics, the word ``constraint" is the most familiar; hence, it is employed here.
The nature of constraints in materials was not known at that time. Nowadays, we know that the substance of constraint is an energy barrier of any kind. In the introductory course of thermodynamics, the word constraint is used to mean macroscopic walls like a rigid wall of a container. The wall is visible, and hence the meaning is clear. In contrast, since the energy barrier in materials is invisible, the role of the energy barrier in thermodynamics is clouded. However, by considering the physical reality of the macroscopic wall of a container, which may be made of metal, we will find that the substance of the macroscopic wall is another kind of energy barrier built inside the metal. 
The height $E_{b}$ of any energy barrier is finite. This determines the relaxation time $\tau$, as 
\begin{equation}
\tau = \tau_{0} \exp\left( \frac{E_{b}}{k_{\rm B}T} \right),
\label{eq:RelaxationTime}
\end{equation}
where $\tau_{0}$ is the inverse of the attempt frequency of the corresponding motion, and $k_{\rm B}$ is Boltzmann's constant. 
A rigid wall is a mathematical tool to raise $E_{b}$ and $\tau$ to infinity to simplify the analysis. 
\begin{Definition}{\rm (Constraint)}
A constraint $\xi_{j}$ is an object to inhibit the change in $j$th-type motions of the constituting particles of a system.
\label{def:constraint}
\end{Definition} 
A constraint $\xi_{j}$ is substantiated by the energy barrier $E_{b,j}$ for $j$th-type motions. Since in the real world the relaxation time $\tau_{j}$ is always finite due to the finite value of barrier height $E_{b,j}$, the $j$th constraint works only in a timescale shorter than $\tau_{j}$. 

A dynamical variable $X(t)$ is subject to a certain constraint $\xi_{j}$. 
The constraint $\xi_{j}$ on $j$th-type of motions means the imposition of a range of permissible changes in $X(t)$. 
\begin{Definition}{\rm (State variable)}
If the time average $\overline{X(t)}$ has a definite value when the system is in equilibrium, the state variable of $j$th type, $X_{j}$, is defined by 
\begin{equation}
X_{j} = \frac{1}{t_{0}} \int_{\xi_{j}} X(t) dt,
\label{eq:jth-SV}
\end{equation}
where $t_{0}$ is the period of time average. \label{def:SV}
\end{Definition} 
By the definite value, we mean that the time-averaged value is independent of the time period $t_{0}$, in which the time average is taken. For example, for a gas in a box with volume $V$, the constraint is the wall of the box. The position, $x(t)$, of a gas molecule is constrained within the wall. In equilibrium, the molecule visits everywhere in the box with equal probability. The time average of the molecule's position, $\overline{x(t)}$, is indeterminate. Thus, $\overline{x}$ is not a state variable.
Now consider the spatial distribution function, $\delta(x', x(t))$, of a molecule at $x'$. The integration of this function over the inside of the box
\begin{equation}
X(t) = \int_{V} \delta(x', x(t)) dx',
\label{eq:int-distribution}
\end{equation}
is interpreted as the density-weighted volume. When the distribution is uniform in space, which is obtained in equilibrium, the time-averaged value $\overline{X(t)}$ gives the volume $V$. The time-averaged value $\overline{X(t)}$ turns to a state variable $V$. The non-vanishing property after time average is the most important property of state variables, and here it is called the {\em time invariance} against time averaging: it is different from the usual one, i.e., the constant in the equation of motion. Or we may call it the invariance against noise. This invariance is exploited to identify state variables in the following.

A thermodynamic system consists of many constraints $\{ \xi_{j} \}$ (Ref.~\cite{Gyftopoulos}, p.~13). Indeed, the substances that characterize the structure of a system are constraints. The wall of a container determines the shape of the gas in the container. At a fixed temperature, only one independent state variable $V$ exists. When a mobile wall is inserted into this container, the structure of the whole system is specified by two volumes $V_{1}$ and $V_{2}$, separated by the internal wall. The state variables are these two volumes $V_{1}$ and $V_{2}$. This partitioning can be continued to any number of compartments. We find that one new state variable is created each time a constraint is inserted. This one-to-one correspondence between constraint and state variable was clarified by Reiss \cite{Reiss}. The structure of a system is fully specified by a set of constraints $\{ \xi_{j} \}$. 

The heuristic finding of this one-to-one correspondence between constraint and state variable is proven rigorously from the second law of thermodynamics.
The second law of thermodynamics is exceptional in physics laws in that it has many different ways of expression, namely, 21 different ways \cite{Capek}. The most recent one, given by Gyftopoulos and Berreta, reads as follows:

\noindent
{\bf The second law of thermodynamics}

{\em There is one and only one equilibrium state for a set of constraints $\{ \xi_{j} \}$ and a given $U$.} {\rm (Ref.~\cite{Gyftopoulos}, p.~63)} 

\noindent
This guarantees that the state variable $X_{j}$ is uniquely determined for a constraint $\xi_{j}$ ({\em uniqueness}). As stated in Definition \ref{def:SV}, the uniqueness is the definiteness that the state variables must satisfy. For example, the time averaged positions of atoms in a gas do not have a unique value; hence, they cannot be state variables.
Furthermore, this expression of the second law also claims that equilibrium is uniquely specified by a set of constraints $\{ \xi_{j} \}$ ({\em completeness}).

By defining state variables rigorously, we find that the equilibrium positions of atoms in solids are state variables. An atom $j$ in a solid is so constrained that it can move only within its unit cells $\xi_{j}$. It has a unique value $\bar{\bf R}_{j}$ in equilibrium. Therefore, we conclude
\begin{Corollary}{\rm (State variables of solids)}
State variables of a solid are the time-averaged positions, $\bar{\bf R}_{j}$, of all atoms that comprise the solid.
\label{col:atom-position}
\end{Corollary} 
The full set $\{ \bar{\bf R}_{j} \}$ of a solid uniquely determines the structure of a solid. This means that completeness is satisfied by this set. The internal energy of a solid is expressed as a function of $\{ \bar{\bf R}_{j} \}$ in addition to a trivial variable $T$,
\begin{equation}
U = U(T, V, \{ \bar{\bf R}_{j} \}).
\label{eq:internalU}
\end{equation}
Besides $T$, there are $3N_{\rm at}$ state variables in a solid, where $N_{\rm at}$ is the number of atoms in the solid. This is the most contrasting thermodynamic property of solids compared to that of gases, in which only volume $V$ is a state variable. Up to here, no assumption has been made about the periodicity or symmetry of solids. Hence, Corollary \ref{col:atom-position} also applies to glasses.

The functional relation of Eq.~(\ref{eq:internalU}) is taken for granted for today's microscopic theory of solids, although $\{ \bar{\bf R}_{j} \}$ were not acknowledged as state variables. A displacement of only one atom to an interstitial position modifies the internal energy $U$. Interstitial positions are also equilibrium positions in the present context, although they are often regarded as metastable states in materials science. However, a metastable state is also an equilibrium state within the timescale where the interstitial atom stays at that position. Therefore, as in Eq.~(\ref{eq:internalU}), $3N_{\rm at}$ equilibrium positions are independent variables. When a solid is transformed into a liquid, the atom positions lose their role as state variables because the time average of atom position becomes indeterminate.
A block of ice in a room is thermodynamically stable in a few minutes, and we can measure its thermodynamic properties on this time scale. Equilibrium atom positions $\bar{\bf R}_{j}$ of the ice are state variables. We can regard the ice as an equilibrium state. However, it will thaw out in a day. The ice state must be considered as a nonequilibrium state on this timescale. Equilibrium atom positions $\bar{\bf R}_{j}$ become indeterminate. After melting, atom positions disappear from the arguments of Eq.~(\ref{eq:internalU}).

By the arguments made up to here, it is concluded that glasses are in equilibrium states, similar to other solids. They do not change their structures within their own relaxation times. Each state of the glass on the quasi-static path $b$ in Fig.~\ref{fig:CTexperiment} is fully specified by state variables $\{ \bar{\bf R}_{j} \}$ through Eq.~(\ref{eq:internalU}). This conclusion is quite different from the traditional view of nonequilibrium character of glass. Over a century, it has been believed that glasses are in nonequilibrium states. However, it is noted that previous studies cannot be rigorous proof for this: it is illogical to prove nonequilibrium when the rigorous definition of equilibrium is not known. Previous studies constructed their theories by placing the nonequilibrium character of glass at the starting point, while this character itself was taken for granted. The main reasoning behind this characterization is that the current properties of a glass cannot be specified by $T$ and $V$ solely \cite{note-third-law}. However, as already pointed out, there is no proof that only $T$ and $V$ are independent state variables for solids. Rather, it is shown that $\{ \bar{\bf R}_{j} \}$ are needed for complete thermodynamic description of a solid.
Now, it is demonstrated that the properties of glasses are better described by adapting this equilibrium view: these include a free-energy behavior \cite{Shirai20-GlassState,Shirai21-GlassHysteresis}, the jump behavior of specific heat at the glass transition \cite{Shirai22-SH,Shirai22-Silica}, and the large activation energies of the glass transition \cite{Shirai21-ActEnergy}.

Lastly, we comment on two words, ``macroscopic" and ``few numbers", when discussing state variables. It is commonly said that state variables are macroscopic quantities, which cannot accept atom positions as state variables. However, how large is macroscopic and how small is microscopic are ambiguous. Often, we distinguish them on a human scale. This manner is subjective, which is not suitable for describing physics. Atom positions are microscopic on a human scale. However, we can observe it, for example, using the X-ray diffraction method. The diffraction pattern exposed on a film is macroscopic. If we do not know the mechanism of the X-ray diffraction method, we might recognize that the crystal structure is characterized by a macroscopic entity.
Also, it is widely believed that the number of state variables must be small, which cannot accept the Avogadro's number of variables. However, again, it is ambiguous how big a large number is and how small a small number is. Infinity is a large number. However, there are many infinities. For example, the number of integers is infinity, while it is quite smaller than the number of real numbers. By definition, an average $\bar{\bf R}$ presumes the existence of a large number of microscopic states ${\bf R}(t)$.

\subsection{Order parameters of glass}
\label{sec:OP}
The immediate consequence of the preceding argument is that OPs are state variables. An OP is a physical property that does not vanish by time averaging in equilibrium. It retains a constant value as long as an equilibrium does not break, namely, the time invariance. This property is fully compatible with Definition \ref{def:SV}. Therefore, from Corollary \ref{col:atom-position}, we conclude the following.
\begin{Corollary}{\rm (Order parameters of solids)}
Order parameters of a solid are state variables of that solid.
\label{col:OPsolid}
\end{Corollary} 
A familiar example of OP is magnetization in magnetic materials. Above the Curie temperature $T_{c}$, the orientation of local moments ${\bf m}_{j}(t)$ is random, and hence its time average vanishes. Below $T_{c}$, the time average does not kill the nonzero component of the local moment $\bar{\bf m}_{j}$. Hence, the time-averaged moment $\bar{\bf m}_{j}$ can be a state variable. 

The total magnetization ${\bf M}$ is considered the OP for the ferromagnetic case. However, vanishing total moment, ${\bf M}=0$, does not necessarily mean lack of order.
For ferromagnetic crystals, the time-averaged local moments $\bar{\bf m}_{j}$ are all the same, $\bar{\bf m}$, yielding the nonzero total magnetization ${\bf M} = \sum_{j} \bar{\bf m}_{j} = N_{\rm at} \bar{\bf m}$. The number of OPs, $n_{\rm OP}$, is one.
For anti-ferromagnetic crystals, there are two types of sites ($k=1$ and $2$). There is the same number for each site, $N_{\rm at,k}$. In each type of sites, the time-averaged local moments $\bar{\bf m}_{j}$ all are the same, $\bar{\bf m}_{k,j}=\bar{\bf m}_{k}$, yielding the subtotal ${\bf M}_{k} = N_{\rm at,k} \bar{\bf m}_{k}$. The orientations of ${\bf M}_{k}$ of the two groups are anti-parallel so that the total moment ${\bf M}$ vanishes. In this case, we can take ${\bf M}_{1}$ or ${\bf M}_{2}$ as the single OP with an additional constraint ${\bf M}_{2} = -{\bf M}_{1}$. Let this constraint be an additional OP, then $n_{\rm OP}=2$. We can continue creating different kinds of magnetic order by increasing the number of magnetic sites $k$ to an arbitrary number. There is no macroscopic value in the total magnetization ${\bf M}$, but an OP still exists, and there may be more OPs.
Spin glass is a magnetically ordered state in the limit of an infinite $n_{\rm OP}$. Each local moment has a random orientation but a non-vanishing averaged value $\bar{\bf m}_{j}$. This is a contrasting property to paramagnetism, where all the time-average local moments vanish, i.e., $\bar{\bf m}_{j}=0$. A local moment $\bar{\bf m}_{j}$ is a microscopic quantity on a human scale but has the time-invariance property, and therefore is a state variable and also an OP.

Overwhelming magnetism is carried by electron spins, although spin freedoms are not considered here. If the readers do not want to be bothered with extra freedoms, they may replace the above examples of magnetism with ferroelectricity (or antiferroelectricity). This can be done simply by reinterpreting the local moment $m_{j}$ as the local dipole moment. The dipole moment is presented by atom displacement $\bar{\bf u}$, as $\bar{\bf m}_{j} = q_{j} \bar{\bf u}_{j}$, where $q_{j}$ is an effective charge. In this manner, all quantities can be represented by atom positions $\bar{\bf R}_{j}$, so all arguments up to here can be directly applied. Similar to magnetism, we can create as many OPs as desired. The limiting case $n_{\rm OP} = 3 N_{\rm at}$ is ``dipolar glass", in analogy with spin glass. Indeed, this name is used for some kind of asymmetric molecular crystals such as rubidium cyanide \cite{Shimada86}. Such molecular crystals are known to have non-vanishing entropy at $T=0$ (in Ref.~\cite{Fowler-Guggenheim}, \S 532). 

An apparent OP presenting the solid structure is an X-ray diffraction pattern (listed in Table 1 of Ref.~\cite{Anderson84}). They appear when a liquid is crystallized and they change when a phase transition occurs between different structures. However, it is rare to acknowledge X-ray diffraction patterns as OP, probably because it is too obvious, so it is difficult to find merit in treating them as OPs. 
The position of atoms appearing in the X-ray diffraction is the time-averaged position $\bar{\bf R}_{j}$ but not instantaneous position ${\bf R}_{j}(t)$. (An interesting historical debate is written in Ref.~\cite{Kittel-ISSP8}, p.~641). Therefore, the X-ray diffraction patterns are qualified as OP of the crystal.

Certainly, X-ray diffraction patterns are obtained because of crystal periodicity. As a result, it is liable to consider OP as the periodicity in this case. This is also why it is difficult to find OP in amorphous materials \cite{note3}.
Although periodicity is a necessary condition for diffraction, the condition of the definiteness of $\bar{\bf R}_{j}$ precedes the presence of periodicity. In the case of solids, regardless of periodicity, the same-time position-position correlations $\langle {\bf R}_{k}(t), {\bf R}_{j}(t) \rangle_{t}$ have definite values, $\bar{\bf R}_{k} \otimes \bar{\bf R}_{j}$, but not for liquids and gases. Therefore, $\{ \bar{\bf R}_{j} \}$ should be OP as well as state variables, and this is true for amorphous solids too.
One might complain that there is no method to observe correlation functions $\langle {\bf R}_{k}(t), {\bf R}_{j}(t) \rangle_{t}$ of individual pairs of atoms and argue that non-observable quantities cannot be OPs. 
The X-ray diffraction method extracts spatial periodicity from the correlation function $\langle {\bf R}_{k}(t), {\bf R}_{j}(t) \rangle_{t}$ by averaging over all the atoms. Typically, experimentalists use the powder X-ray diffraction method among many variants. Using this method, the periodicity of the three-dimensional space is averaged to reduce to one-dimensional information. If the powder diffraction method were the only available experiment, only the orientational average of diffraction peaks would correspond to OP. This is tantamount to say that the definition of OP changes as our technology is developed. This is undesirable for describing the principles of physics. The measurability in diffraction methods is merely a problem of currently available experimental techniques and should be irrelevant to the definition of state variables. Direct imaging methods, such as transmission electron microscopy, are now available. Although their utilization is still limited, giving one counterexample is enough for denying the impossibility of direct observation of microscopic structures.

For the glass transition, in old theories, the OP $z$ is interpreted to present a nonequilibrium parameter, as described in Introduction. It is considered that a glass-forming liquid freezes in the cooling process before reaching equilibrium. Thus, $z$ is frozen at nonzero value $z=z_{0}$. The value $z_{0}$ is determined by the condition of minimum free energy $G$, according to Eq.~(\ref{eq:dgdz-zero}). Thus, $z_{0}$ is considered not to be an independent variable, as $z_{0}=z_{0}(T, p)$ \cite{Landau-SP}. 
In the new theory, the OPs are equilibrium positions, $\bar{\bf R}_{j}$, for each frozen state. Furthermore, these OPs are independent of $T$ and $p$. In the new theory, time averaging must be performed within the constraints $\{ \xi_{j} \}$. With these constraints, the position of $j$th atom, ${\bf R}_{j}(t) = \bar{\bf R}_{j}+ {\bf u}_{j}(t)$ can vary only in the vicinity of $\bar{\bf R}_{j}$. Hence, the accurate form of Eq.~(\ref{eq:dgdz-zero}) must be interpreted as follows:
\begin{equation}
\left( \frac{\partial G}{\partial {\bf u}_{j} } \right)_{T,p, \{ \bar{\bf R}_{j}\} }=0. 
\label{eq:dgdz1-zero}
\end{equation}
It has been known that there are infinite numbers of local minima in the energy functional $E=E(\{ {\bf R}_{j}\})$ in a solid \cite{Payne92,Oganov06}. This fact can be understood by Eq.~(\ref{eq:dgdz1-zero}).
There is no surprise to find infinite equilibrium configurations in a single glass. All such states can be described by thermodynamic methods.

\section{Properties of the glass order parameter}
\label{sec:discussion}

\subsection{Plurality and arbitrariness}
\label{sec:Plurality}
In this section, some OP features in glass are discussed.
The first one is arbitrariness in choosing the specific form of OPs. 
The definition of OP in glasses varies in the literature. Some properties of glasses are often used for the OP. In Sec.~\ref{sec:OP}, on the other hand, we have demonstrated that OPs are the equilibrium positions of the atoms. Here, the gap between the present conclusion and others' view is explained.
The most widely-used OP in the glass literature may be the fictive temperature, which was phenomenologically introduced by Tool about a century ago \cite{Tool31,Tool46}. He defined the fictive temperature $T_{f}$ as the temperature at which the glass would find itself in equilibrium if brought there quickly from its current state. Later, his formula was better treated in a thermodynamic framework by Davies and Jones \cite{Davies53a}. 
The idea of introducing the fictive temperature is to facilitate thermodynamic treatments by bypassing the relaxation process. The equilibrium condition (\ref{eq:dgdz-zero}) is equivalent to $A(T,p,z)=0$. Hence, the fictive temperature is defined by:
\begin{equation}
A(T_{f},p,z)=0.
\label{eq:fictiveT}
\end{equation}
Thus, $T_{f}$ is a function of $p$ and $z$. In this manner, the fictive temperature can be qualified as a state variable and also an OP, provided that there is a unique $T_{f}$ satisfying Eq.~(\ref{eq:fictiveT}) for a given $z$.

Equation (\ref{eq:fictiveT}) is not too useful unless the concrete form $A(T, p, z)$ is known. 
The response of glass to a given external influence exhibits non-linearity and non-exponential time response \cite{Hodge94}. By acknowledging that the value of fictive temperature depends on the measured property $P$, Narayanaswamy defined $T_{f}$ by the response function $M$
\begin{equation}
M(t, \Delta T) = \frac{P_{2}(t)-P_{2}(\infty)}{P(0)-P_{2}(\infty)} = \frac{T_{f}-T_{2}}{\Delta T}, 
\label{eq:fictiveT}
\end{equation}
for a step temperature change $\Delta T=T_{2}-T_{1}$ \cite{Narayanaswamy71}. Here, $P_{2}(t)$ stands for the time-dependent $P$ after changing $T$ from $T_{1}$ to $T_{2}$. Using $M(t)$, the current value $P(t)$ is calculated by time convolution of the history of thermal treatment $T(t')$, where $t'\leq t$. Further approximations are needed to work on the formula, but it is not the subject of this study. Equation (\ref{eq:fictiveT}) shows that any property of glass can be used as an OP through the fictive temperature. In fact, enthalpy, thermal expansivity, and refractive index represent OPs, with different $T_{f}$ \cite{Ritland54-56,Kovacs79,Hutchinson76}. 
However, it is noted that, from the present view, there is a problem in the previous usage of $T_{f}$. The problem is the assumption for the final state of the relaxation process. Many researchers believe that, in the transition region, the glass state is metastable. In stead, supercooled liquids are considered as equilibrium, and the glass state is expected to be eventually transformed into a liquid state over a long period \cite{Davies53a}. This assumption is not approved by the energetic argument made by the author \cite{Shirai20-GlassState}, which shows that the free energy of glass is lower than that of the parent supercooled liquid. The supercooled liquid state cannot be restored at $T<T_{g}$.

The plurality of OPs mentioned above is evident from the present theory. Since the properties of solids are determined by their structures, any physical property $P_{k}$ can be taken as OP,
\begin{equation}
P_{k} = P_{k}( \{ \bar{\bf R}_{j} \} ).
\label{eq:Property}
\end{equation}
There are $N_{\rm at}$ independent properties $P_{k}$. Among them, any set $\{ P_{k} \}$ can be used for a set of OPs.
The fictive temperature, $T_{f,k}$, associated with the property $P_{k}$ is determined by a set of constraint, $\{ \xi_{j} \}$. As described in Sec.~\ref{sec:equilibrium}, every constraint has its own relaxation time $\tau_{j}$. Even if the volume is fixed, a relaxation occurs in the enthalpy. Accordingly, the relaxation times of volume expansivity and of enthalpy are different. The refractive index is affected by the inhomogeneity of the sample, whose relaxation process is different from other quantities.

\subsection{Randomness and order}
\label{sec:randomness}

The second issue is the meaning of disorder in thermodynamics. This issue was already discussed in Sec.~\ref{sec:OP}. However, it is useful to investigate it from a different view point of the notion {\it missing information}, which may be a better interpretation for entropy than the conventional interpretation, i.e., the degree of disorder is \cite{Ben-Naim}. In solid state physics, disorder means a break of periodicity. This traditional interpretation penetrates so deeply into our minds that we are liable to understand that the periodicity is akin to the order state in thermodynamics too. However, by developing the material research, many solids with non-periodic structures have been discovered, including incommensurate compounds and quasicrystals. It is difficult to say whether they are ordered states or not. Outside solid state physics, the traditional interpretation of disorder for entropy was already criticized \cite{Ben-Naim,Rosenkrantz83, Grandy,Denbigh89,Styer00}. Thermodynamics is a universal theory whose laws must remain independent of the details of material structures.

\begin{figure}[htbp]
    \centering
    \includegraphics[width=58 mm, bb=0 0 620 500]{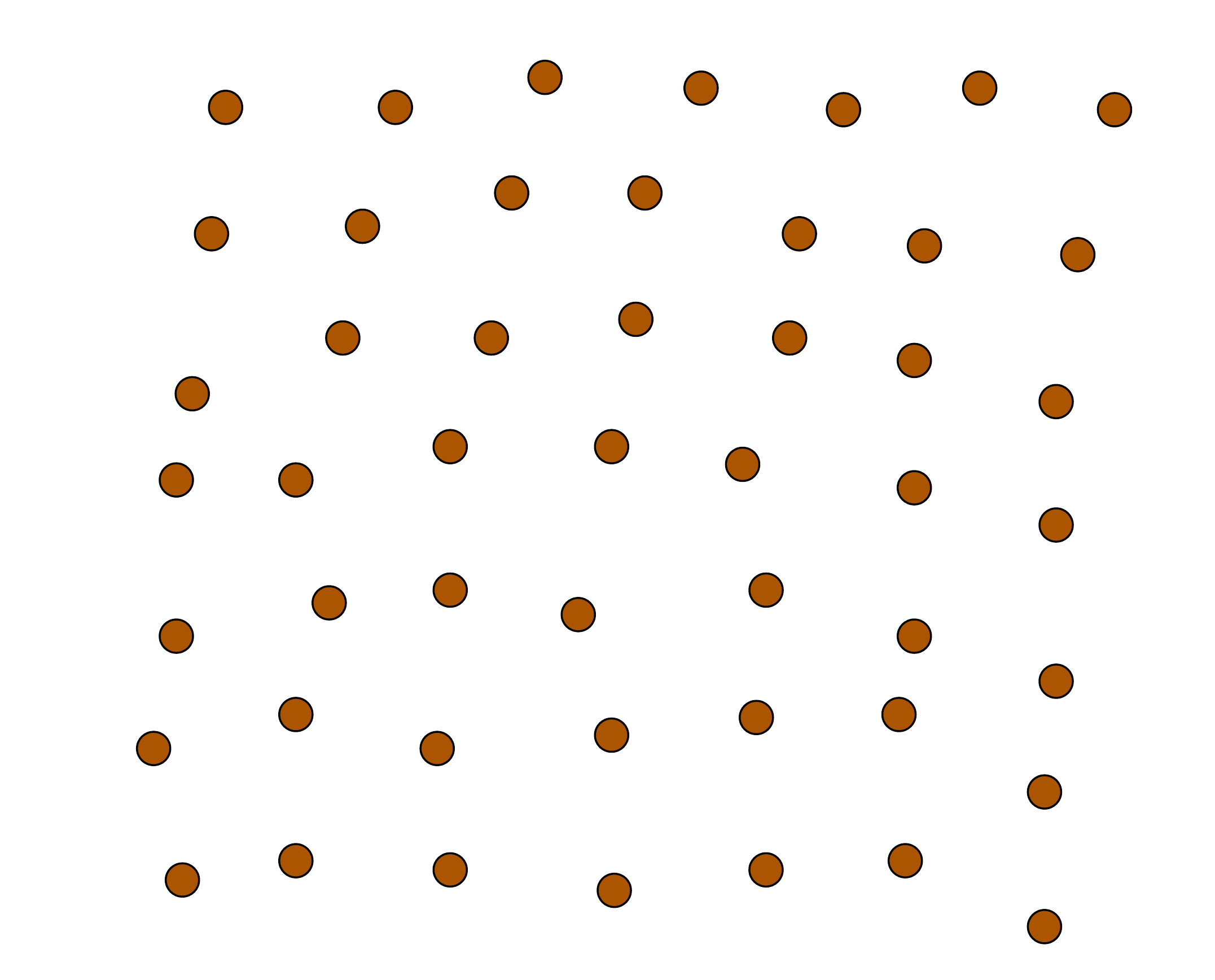} 
    \hspace{5 mm}
    \includegraphics[width=58 mm, bb=0 0 620 500]{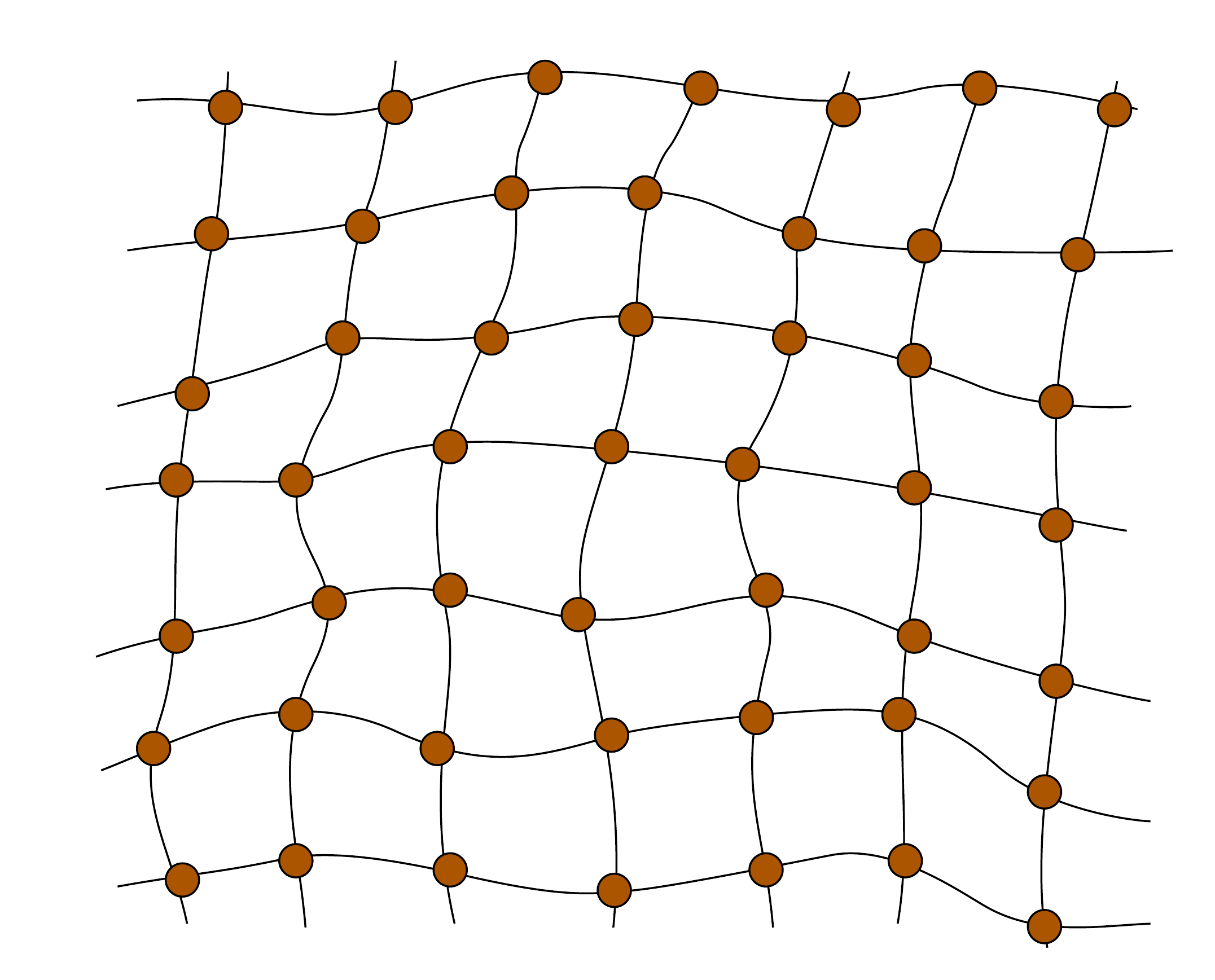} 
  \caption{(a) Random arrangement of atoms. This figure is the same as (b), except the lines are hidden.
  } \label{fig:array3}
\end{figure}

The meaning of order/disorder in thermodynamics was seriously studied in information theory, from which the notion of missing information emerged \cite{Jaynes57-all,Tribus71,Wehrl78,Jaynes79,Zurek99,Haar-Thermostat,Ben-Naim}. If we look at the number 3.14159$\dots$, it looks like a random number but yet conveys useful information \cite{Chaitin-SAs}. An English text would appear to be just a series of random letters to someone who does not know English. If we do not know the meaning of the arrangement of nucleotides in DNA, they seem merely a random array. These examples urge us to reflect on the definition of randomness. In information theory, where there is no idea of crystal symmetry, the missing information has generality.
An atom arrangement is shown in Fig.~\ref{fig:array3}(a). There is no periodicity in the atom arrangement. However, this figure was drawn by removing the lines in Fig.~\ref{fig:array3}(b), in which a square lattice is drawn in a distorted space. In a square lattice in the Cartesian coordinates, each atom is identified by the lattice address $(m,n)$. The information of the address of atoms is not missed by deforming the space. Topologically, the two figures are equivalent.
The geometrical arrangement of homes in our town does not have periodicity. However, a unique address is assigned to each home. A mail carrier can deliver postal materials to any home once he retains the town map. However, if residents move frequently, the mail carrier cannot deliver the items to the residents correctly. Information on addresses is missing. In order to correctly deliver the postal items, the mail carrier must ensure that the address of today is the same as yesterday. As far as the word ``apple" is always associated with the real object apple, this word conveys useful information. If everyone uses it differently, this word does not make sense. The common property of information in the above examples is the time invariance: that is, it always gives the same value at any time it is measured. 
This requirement of the time invariance is the same as for state variables, as stated in Definition \ref{def:SV}.
\begin{Corollary}{\rm (Information)}
Information in thermodynamics means values of state variables.
\label{col:info}
\end{Corollary} 
State variables (and OPs) play the role of language, conveying the information of a material \cite{note4}. 

Sometimes, the notion of missing information is misconstrued. There is a long debate about the nature of information when the meaning of entropy is investigated \cite{Jaynes79,Denbigh81}. It is said that there are two approaches to statistical thermodynamics: objective and subjective information (in Ref.~\cite{Callen}, p.~385). The issue is, who is the subject of missing information \cite{Denbigh81}?
The entropy value of DNA is a property of that DNA and should not be changed if we learn the meaning of the arrangement of nucleotides. The essential matter is not the missing of {\em our} information but the time invariance of material properties. A specific arrangement of nucleotides will have the same result whether we observe it today or tomorrow. The arrangement of nucleotides is thus state variables and, thereby, OPs. 

There are often redundancies in a full set of state variables $\{ \bar{\bf R}_{j} \}$. Crystal symmetry yields redundancy. In the case of ferromagnetic materials, individual moments $\bar{\bf m}_{j}$ are all the same. Hence, only one OP of the total magnetization $\bar{\bf M}$ is a useful state variable. In crystals, the periodicity significantly reduces the independent state variables among $\{ \bar{\bf R}_{j} \}$. Rotational symmetries impose further restrictions on the remaining state variables. In this manner, the role of crystal symmetries in thermodynamics reduces the number of independent OPs but does not change the essential nature of state variables, namely, the time invariance. 

\subsection{Prigogine-Defay ratio}
\label{sec:PDratio}

Interesting evidence for the plurality of OPs in glasses stems from the inequality relationship for the Prigogine-Defay (PD) ratio ${\it \Pi} \equiv \Delta C_{p} \Delta \kappa / TV (\Delta \alpha)^{2}$. Here, $\Delta C_{p}$, $\Delta \kappa$ and $\Delta \alpha$ are the jumps in specific heat, in compressibility, and in thermal expansion coefficient, respectively, at the transition \cite{Prigogine54}. In standard phase transition theory for crystals, the same quantity is defined for the second-order transition and is proven as ${\it \Pi} = 1$ \cite{Pippard}.
For the glass transition, the overwhelming experimental data indicate inequality, \cite{Davies53a,O'Reilly62,Moynihan76a,Moynihan81}
\begin{equation}
{\it \Pi}  > 1.
\label{eq:PDrelation}
\end{equation}
Davies and Jones explained this inequality (\ref{eq:PDrelation}) by the presence of more than one OPs \cite{Davies53a}. Later, further proofs were given from various points of view \cite{Goldstein63,Goldstein73,Gupta76,Goldstein75,Lesikar80, Nieuwenhuizen97,Schmelzer06,Tropin12}. 
So far, these proofs are based on the analytical behaviors of thermodynamic functions, leaving the actual substance of OPs untouched. 

From the present theory, it is evident that there is more than one OP. Here, a further significance of the PD ratio is mentioned.
Recently, an attractive physical meaning has been found for this inequality (\ref{eq:PDrelation}) by Shirai {\it et al.} \cite{Shirai22-Silica}. The internal energy $U$ is expressed as a function of the equilibrium positions $\{ \bar{\bf R}_{j} \}$ of all atoms, as shown by Eq.~(\ref{eq:internalU}). This can be broken into 
\begin{equation}
U = E_{\rm st}(T, \{ \bar{\bf R}_{j} \} ) + E_{\rm ph}(T) + E_{\rm te}(T, V),
\label{eq:internal_energy2}
\end{equation}
where $E_{\rm st}$, $E_{\rm ph}$, and $E_{\rm te}$ are the components of the structure (configuration), phonon, and thermal expansion, respectively. 
Correspondingly, the total specific heat is written as
\begin{equation}
C_{p} = C_{\rm st}(T, \{ \bar{\bf R}_{j} \} ) + C_{\rm ph}(T) + C_{\rm te}(T,V).
\label{eq:specific-heat}
\end{equation}
They showed that the PD ratio is rewritten as
\begin{equation}
{\it \Pi} \equiv \frac{ \Delta C_{p}}{\Delta C_{\rm te}}
    = \frac{ 
    \text{(Change in the total energy)} }{ \text{(Contribution to isotropic volume change)} }.
\label{eq:PDratio3}
\end{equation}
This shows that ${\it \Pi}$ is in fact equal to one only when the structural part $\Delta C_{\rm st}$ continuously vanishes at the transition. The equality holds for the second-order phase transition, because there is no discontinuity in the crystal structure.
The glass literature often argues that the glass transition is a kinetic transition because the lack of latent heat is interpreted as no structural change. The fact is that the structural change occurs in the transition region with a finite width $\Delta T_{g}$. This was recently shown by first-principles calculations \cite{Shirai22-SH,Shirai22-Silica}. These authors showed that the structural part $\Delta C_{\rm st}$ dominates in $\Delta C_{p}$. 
Therefore, the glass transition can be essentially classified as the first-order transition. The lack of latent heat is due to broadening the transition temperature $T_{g}$. Latent heat $Q_{l}$ emerges only due to the occurrence of transition at a discrete $T$. Pippard points out that what matters is whether there is an energy difference before and after the transition, not whether latent heat is observed (in Ref.~\cite{Pippard85}, Chap.~8). Defects and impurities destroy the long-range order. This breaks latent heat at the discrete temperature $T_{c}$, which turns to sensible heat in a transition range with a width $\Delta T_{c}$, resulting in a substantial increase in $C_{p}$ in this region.

\section{Conclusions}
\label{sec:conclusion}
We have shown that the notions of the order parameter, information, and state variable are the same in thermodynamics. They have a common property of the time invariance against time averaging. Their measurement returns the same value every time it is performed once equilibrium was established.
Based on the time-invariance property, we conclude that the glass OPs are the equilibrium positions $\{ \bar{\bf R}_{j} \}$ of all the atoms that constitute that glass. Any glass property that is determined by the structure $\{ \bar{\bf R}_{j} \}$ can be an OP. Crystal symmetry brings about redundancy in the OPs and reduces the number of independent OPs, whereas all OPs remain independent variables of a glass. The often-claimed association of orders with periodicity is disproven by abundant examples of well-ordered materials, such as DNA. The orders in thermodynamics instead mean the time invariance.

The above conclusion is derived from the rigorous definition of equilibrium, which used only the second law of thermodynamics. 
There is no assumption that the ordered state is the lowest-energy state, as is often claimed. Indeed, this assumption is inconsistent with experimental facts.
The never-proven speculation ``an ordered state will be obtained if we wait a long time" has been eliminated from the theory. All equilibria can be retained only within the timescale in which a given set of constraints is maintained. Accordingly, OPs, state variables, and information about states can only remain meaningful within that timescale.

Usefulness of the present theory in current research interests is that it provides firm grounds on the machine leaning of glasses \cite{Bapst20,Minamitani23} and phase change materials \cite{Sosso19}. The theory guarantees that the attempt to specify the current properties of these materials by static parameters in their language must in principle succeed.

\begin{acknowledgments}
The author thanks H. Yoshino for the valuable discussion on the Edward-Anderson model. 
We also thank Enago (www.enago.jp) for the English language review.
We received financial support from the Research Program of ``Five-star Alliance" in ``NJRC Mater.~\& Dev." (2020G1SA028) from the Ministry of Education, Culture, Sports, Science and Technology of Japan (MEXT).
\end{acknowledgments}





\bibliography{thermo-refs, glass-refs, added-refs}

\begin{thebibliography}{101}
\bibitem[{not({\natexlab{a}})}]{note1}
\bibinfo{note}{The random first-order transition (RFOT) theory has been
  developed for glass transition in analogy with the Edward-Anderson (EA) model
  in spin glass \cite{Kirkpatrick89,Mezard12}. The original EA theory
  \cite{Edwards75} uses a mathematical tool of replica ensemble. Physical
  reality of such a model requires further interpretation, and hence it is out
  of the present scope.}
\bibitem[{not({\natexlab{b}})}]{note2}
\bibinfo{note}{Now, the traditional view that the glass state is a
  nonequilibrium state is subjected to serious criticisms. Many pieces of
  compelling evidence supporting the stability of silica glass exist. Silica
  glass in geology retains its structure more than one million years
  \cite{Berthier16}, while normal metals degrade only in a thousand years.
  Recent experiments do not support the predicted long-term changes
  \cite{Bruning92,Hecksher08, Zhao13,Ponga15}. Theoretically, it is shown that
  the free energy of glass is lower than that of supercooled liquid at $T <
  T_{g}$ \cite{Shirai20-GlassState}. In first-principles calculations, there is
  no sign for the mechanical instability of structures.}
\bibitem[{not({\natexlab{d}})}]{note3}
\bibinfo{note}{In this paper, the terms of amorphous and glass are used as the
  same meaning. Conventionally, they are considered as distinct classes of
  materials in terms of the forming process. In this study, processes are not
  our concern.}
\bibitem[{not({\natexlab{e}})}]{note4}
\bibinfo{note}{We note a risk of careless use of information. Making too much
  analogy to information theory will fall in endless arguments on the concept
  of information \cite{Jauch03,Beauragard03,Denbigh81,Jaynes79}. To avoid this,
  we restrict the usage of information to the information of material
  properties only. The above arguments on numbers and languages are cited only
  because of usefulness of understanding randomness.}
\end{thebibliography}



\end{document}